# SOURCE CODE ANALYSIS TO REMOVE SECURITY VULNERABILITIES IN JAVA SOCKET PROGRAMS: A CASE STUDY

Natarajan Meghanathan

Jackson State University, 1400 Lynch St, Jackson, MS, USA
natarajan.meghanathan@jsums.edu

## ABSTRACT

*This paper presents the source code analysis of a file reader server socket program (connection-oriented sockets) developed in Java, to illustrate the identification, impact analysis and solutions to remove five important software security vulnerabilities, which if left unattended could severely impact the server running the software and also the network hosting the server. The five vulnerabilities we study in this paper are: (1) Resource Injection, (2) Path Manipulation, (3) System Information Leak, (4) Denial of Service and (5) Unreleased Resource vulnerabilities. We analyze the reason why each of these vulnerabilities occur in the file reader server socket program, discuss the impact of leaving them unattended in the program, and propose solutions to remove each of these vulnerabilities from the program. We also analyze any potential performance tradeoffs (such as increase in code size and loss of features) that could arise while incorporating the proposed solutions on the server program. The proposed solutions are very generic in nature, and can be suitably modified to correct any such vulnerabilities in software developed in any other programming language. We use the Fortify Source Code Analyzer to conduct the source code analysis of the file reader server program, implemented on a Windows XP virtual machine with the standard J2SE v.7 development kit.*

## 1. INTRODUCTION

With the phenomenal growth of the Internet, it is imperative to test for the security of software during its developmental lifecycle and fix the vulnerabilities, if any is found, before deployment. Until recently, security has been often considered as an afterthought, and the bugs are mostly detected post-deployment through user experiences and attacks reported. The bugs are often controlled through patch code (more formally called 'security updates') that is quite often sent to customers via Internet. Sometimes, patch codes developed to fix one bug may often open several new vulnerabilities, which if left unattended, can pose a significant risk for the system (and its associated resources) on which the software is run. It is critical that software be built-in with security features (starting from the requirement analysis stage itself, and implemented with appropriate modules as well as tested with suitable analysis techniques) during its entire development lifecycle.

In this paper, we focus on testing for software security using source code analysis (also invariably referred to as *static code analysis*). Static code analysis refers to examining a piece of code without actually executing it [1]. The technique of evaluating software during its execution is referred to as run-time code analysis (also called dynamic code analysis) [2] – the other commonly used approach to test for software security. While dynamic code analysis is mainly used to test for logical errors and stress test the software by running it in an environment with limited resources, static or source code analysis has been the principal means to evaluate the software with respect to functional, semantic and structural issues including, but not limited to, type checking, style checking, program verification, property checking and bug finding [3]. On the top of these issues, the use of static code analysis to analyze the security aspects of software is gaining prominence. Static code analysis helps to identify the potential threats (vulnerabilities) associated with the software, analyze the complexity involved (in terms of increase in code size, development time, and code run time, etc) and the impact on user experiences in fixing these vulnerabilities through appropriate security controls [4]. Static code analysis also facilitates evaluating the risks involved in only mitigating or just leaving these vulnerabilities unattended – thus, leading to an attack, the consequences of such attacks and the cost of developing security controls and integrating them to the software after the attack has occurred [5]. In addition, static code analysis is also used to analyze the impact of the design and the use of the underlying platform and technologies on the security of the software [6]. For example, programs developed in C/Unix platforms may have buffer overflow vulnerabilities, which are very critical to be identified and mitigated; whereas, buffer overflow vulnerabilities are not an issue for software developed in Java. Software developed for J2EE platforms are strictly forbidden from using a *main* function as the starting point of a program, whereas the *main* function is traditionally considered the starting point of execution of software programs developed in standard J2SE development kits and other high-level programming languages. It would be very time consuming and often ineffective to manually conduct static code analysis on software and analyze the above issues as well as answer pertinent questions related to the security of software. One also needs to have a comprehensive knowledge of possible exploits and their solutions to manually conduct static code analysis.

```
C:\res\source-code-analysis>sourceanalyzer -f fileReaderServer_results.fpr fileR
eaderServer.java

C:\res\source-code-analysis>auditworkbench fileReaderServer_results.fpr

C:\res\source-code-analysis>sourceanalyzer fileReaderServer.java

[C:\res\source-code-analysis]
[06AA5F3F192AB210797CEB614454F9CB : low : J2EE Bad Practices : Sockets : semanti
c ]
fileReaderServer.java(11) : new ServerSocket()

[E4FB26252E0707036EC2C0EC2EE8C30D : low : Denial of Service : semantic ]
fileReaderServer.java(16) : BufferedReader.readLine()

[6A3017B95A13B2CE7CF0515D4C0AEBA9 : low : Denial of Service : semantic ]
fileReaderServer.java(26) : BufferedReader.readLine()

[87622C9095E0F46C95DFF7B4E8545898 : medium : System Information Leak : semantic
]
fileReaderServer.java(40) : Throwable.printStackTrace()

[C60012699B3040DE87BCFFC4FA7BF6E1 : medium : Resource Injection : dataflow ]
fileReaderServer.java(11) :   ->new ServerSocket(0)
    fileReaderServer.java(9) : <=> (serverPortNumber)
    fileReaderServer.java(9) : <->Integer.parseInt(0->return)
    fileReaderServer.java(6) :   ->fileReaderServer.main(0)

[0BC5B81158754D0158921D57CD8DFE2C : medium : Path Manipulation : dataflow ]
fileReaderServer.java(20) :   ->new FileReader(0)
    fileReaderServer.java(16) : <=> (filename)
    fileReaderServer.java(16) : <- BufferedReader.readLine(return)
```

**Figure 1:** Command-line Execution of the Source Code Analyzer on a Java Program and Forwarding the Results to an Audit Workbench Format File

Various automated tools have been recently developed to conduct static code analysis [7][8]. In this paper, we illustrate the use of a very effective tool developed by Fortify Inc., called the Source Code Analyzer (SCA) [9]. The Fortify SCA can be used to conduct static code analysis on C/C++ or Java code and can be run in Windows, Linux or Mac platforms. The SCA can analyze individual program files or entire projects collectively. The analyzer uses criteria that are embedded into a generic rulepack (a set of rules) to analyze programs developed in a specific platform/ language. Users may use these generic rulepacks that come with the SCA or develop their own customized sets of rules.

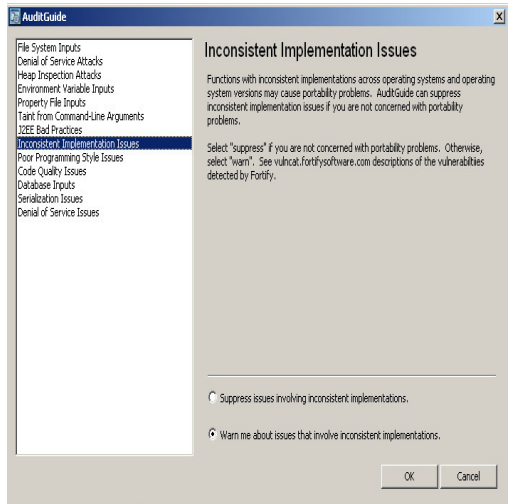 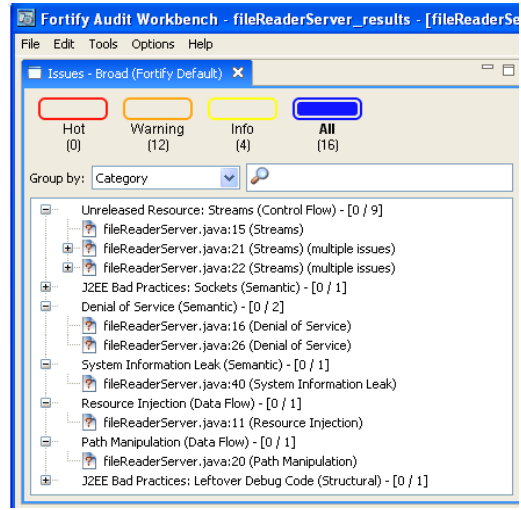

**Figure 2:** Audit Workbench Audit Guide   **Figure 3:** Listing of all the Issues identified

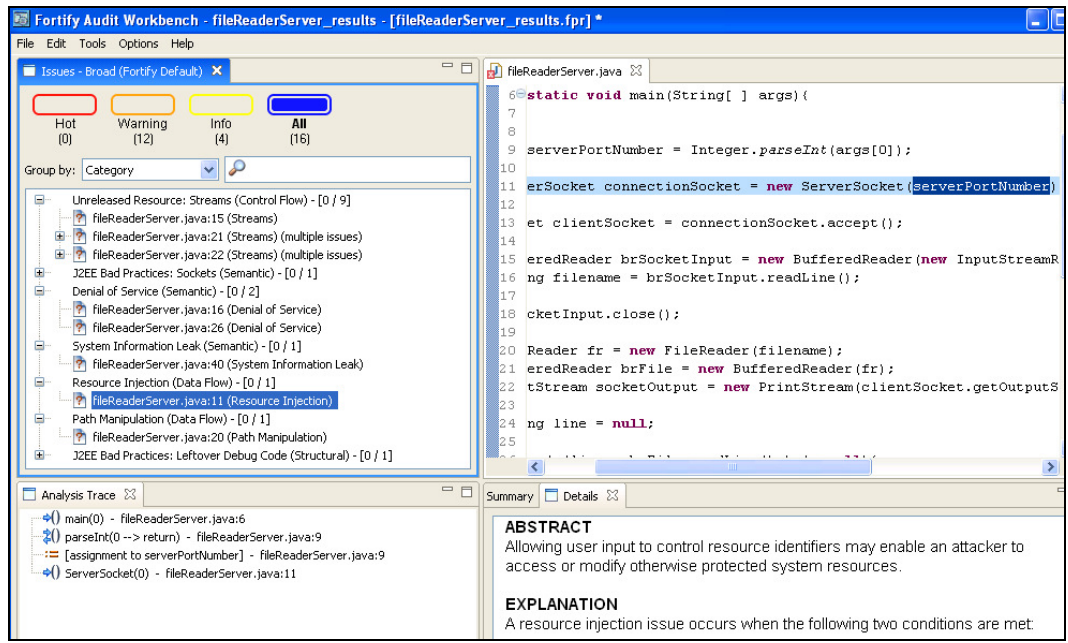

**Figure 4:** Audit Workbench: Issues Panel and Code Editor displaying Details of a Specific Security Issue

The SCA has to be first used in command line (Figure 1) to generate a report, in *.fpr* format (as shown in the first command executed in Figure 1), which can be loaded (second command in Figure 1) into the Audit Workbench utility (screenshot shown in Figure 2), a graphical-user interface utility, included with the Fortify suite of tools. The Workbench interface displays a list of the issues that have been flagged and groups these issues according to their severity (hot, warning, or info). Figure 3 shows a listing of all the issues identified with the file reader server socket program of the case study presented in Section 2. The Workbench includes an editor that can highlight the troublesome code identified to be the source of a particular vulnerability listed in the Issues panel, and also allows users to make changes to the code within the application. Figure 4 shows a comprehensive picture of the Issues panel with the code editor. One significant use of the Workbench utility is that for each generic issue flagged by the analyzer, the utility provides a description of the problem and how it may be averted. If users think that a security issue raised by the analyzer is of no interest to them (i.e. can be left unattended in the code), then the Workbench utility can be set to suppress the raising of the issue in subsequent instantiations of running the analyzer. At any point of time, the suppressed issues can be unchecked and the issues will be raised if found in the code being analyzed at that time. Note that it is important to make sure the source code that is being analyzed compiles without any error prior to running it with the SCA.

```java
1  import java.net.*;
2  import java.io.*;
3
4
5  class fileReaderServer{
6      public static void main(String[ ] args){
7      try{
8
9          int serverPortNumber = Integer.parseInt(args[0]);
10
11         ServerSocket connectionSocket = new ServerSocket(serverPortNumber);
12
13         Socket clientSocket = connectionSocket.accept();
14
15         BufferedReader brSocketInput = new BufferedReader(new InputStreamReader(clientSocket.getInputStream()));
16         String filename = brSocketInput.readLine();
17
18         brSocketInput.close();
19
20         FileReader fr = new FileReader(filename);
21         BufferedReader brFile = new BufferedReader(fr);
22         PrintStream socketOutput = new PrintStream(clientSocket.getOutputStream());
23
24         String line = null;
25
26         while ( (line = brFile.readLine() ) != null){
27
28             socketOutput.println(line);
29
30          }
31
32         brFile.close();
33         fr.close();
34
35         socketOutput.flush( );
36         clientSocket.close( );
37         connectionSocket.close( );
38      }
39      catch(IOException ie){
40         ie.printStackTrace();
41      }
42    }
43  }
```

**Figure 5:** Case Study: Original Java Source Code for the File Reader Server Program

## 2. CASE STUDY ON A CONNECTION-ORIENTED FILE READER SERVER SOCKET PROGRAM IN JAVA

In this section, we present a case study on a file reader server socket program, based on connection-oriented sockets. For simplicity, the server program is considered to serve only one client. The file reader server basically lets a client to read the contents of a file whose name or the path is sent by the client over a socket and the file is locally stored at the server. The server program (whose original source code is shown in Figure 5) works as follows: An object of class *ServerSocket* is instantiated at a port number input by the user. The *ServerSocket* is the class used to open server sockets that wait on a certain port number (publicly known to the clients) for incoming client requests. Once a client contacts the server, the *ServerSocket* is unblocked (through the accept( ) method) and a *Socket* object (in our program the *clientSocket* object) is created as a reference to communicate with the client at the other side. The server waits for the client to send a filename or a pathname through the socket and reads it through a *BufferedReader* object (*brSocket*). Since the server is not sure of the number of characters that would constitute the filename or the pathname, the server uses the *readLine*( ) method of the *BufferedReader* class to read the filename/pathname as a line of characters stored as a String. This String object is directly passed to the *FileReader* constructor to load the file the client wishes to read. The contents of the file are read line-by-line and sent to the client using an object of the *PrintStream* class invoked on the *ClientSocket* object (of class Socket).

We conduct source code analysis of the file reader server socket program (shown in Figure 5) using the Fortify SCA and the output of all the issues identified are shown in Figure 3. Note that the *poor logging practice* warning shown in Figure 3 is due to the use of print statements. We do not bother to remove the print statements and so neglect those warnings. Similarly, we discard the warning message appearing related to *J2EE Bad Practices*: *Sockets*; J2EE standard considers socket-based communication in web applications as prone to error, and permits the use of sockets only for the purpose of communication with legacy systems when no higher-level protocol is available. The Fortify Source Code Analyzer subscribes to the J2EE standards and flags some of the commonly used J2SE features like sockets as something that is vulnerable in the context of security. As mentioned before, the Audit Workbench does provide the flexibility to turn off these flags which do not appear relevant to the programming environment. The goal of the case study is thus to modify the file reader server socket program (and still does what it is intended to do) to the extent that the source code analyzer only outputs warnings corresponding to the poor logging practice and the use of sockets as bad practice, and all the other vulnerabilities and warnings associated with the program are taken care of (i.e., removed).

## 2.1. Resource Injection Vulnerability

The Resource Injection vulnerability (a dataflow issue) arises because of the functionality to let the user (typically the administrator) starting the server program to open the server socket on any port number of his choice. The vulnerability allows user input to control resource identifiers enabling an attacker to access or modify otherwise protected system resources [1]. In the connection server socket program of Figure 5, a Resource Injection vulnerability exist in line 11, wherein the program opens a socket on the port number whose value is directly input by the user. If the server program has privileges to open the socket at any specified port number and the attacker does not have such a privilege on his own, the Resource Injection vulnerability allows an attacker to gain capability to open a socket at the port number of his choice that would not otherwise be permitted. This way, the program could even give the attacker the ability to transmit sensitive information to a third-party server.

We present two solution approaches to completely avoid or at least mitigate the Resource Injection vulnerability: (1) *Use a blacklist or white list*: Blacklisting selectively rejects potentially dangerous characters before further processing the input in a program. However, any such list of unsafe characters is likely to be incomplete and will almost certainly become out of date with time. A white list of allowable characters may be a better strategy because it allows only those inputs whose characters are exclusively listed in the approved set. Due to the difficulty in coming up with a complete list of allowable or non-allowable characters, the approaches of using a blacklist or white list can only mitigate the Resource Injection attack. Nevertheless, if the set of legitimate resource names is too large or too hard to keep track of, it may be more practical to follow a blacklist or white list approach. We will use this approach to remove the Path Manipulation vulnerability in Section 2.2. (2) *Use a level of indirection*: This approach involves creating a list of legitimate resource names that a user is allowed to specify, and only allow the user to select from the list. This approach can help us to completely avoid having Resource Injection vulnerability in the code, because a user cannot directly specify the resource name of his choice and can only chose from what is presented to him. The tradeoff is with the approach of providing a list of port numbers (the resources in our case) to choose from, we are revealing the available port numbers to a user (even though he is constrained only to choose from this list). Note that with the blacklist or white list approach, the user has to merely enter an input of his choice and the program internally processes the input and filters it (thus not revealing information regarding acceptable inputs to the user).

```java
1  import java.net.*;
2  import java.io.*;
3  import java.util.*;
4
5  class fileReaderServer{
6      public static void main(){
7      try{
8
9          int[] availablePortNumbers = {2345, 1234, 8943};
10
11         System.out.println("Choose from the following port numbers to open the socket");
12
13         for (int index = 0; index < availablePortNumbers.length; index++){
14             System.out.println( (index+1)+" --> "+availablePortNumbers[index]);
15         }
16
17         Scanner sc = new Scanner(System.in);
18         int portIndex = sc.nextInt();
19
20         if (portIndex >= 1 && portIndex <= availablePortNumbers.length){
21
22             portIndex--;
23
24             int serverPortNumber = availablePortNumbers[portIndex];
25             ServerSocket connectionSocket = new ServerSocket(serverPortNumber);
26
27             Socket clientSocket = connectionSocket.accept();

53         }
54         else{
55             System.out.println("Error: Wrong selection of port number...");
56         }
57     }
58     catch(IOException ie){
59         ie.printStackTrace();
60     }
61    }
62  }
```

**Figure 6:** Modification to the File Reader Server Program to Remove the Resource Injection Vulnerability (fileReaderServer_1.java)

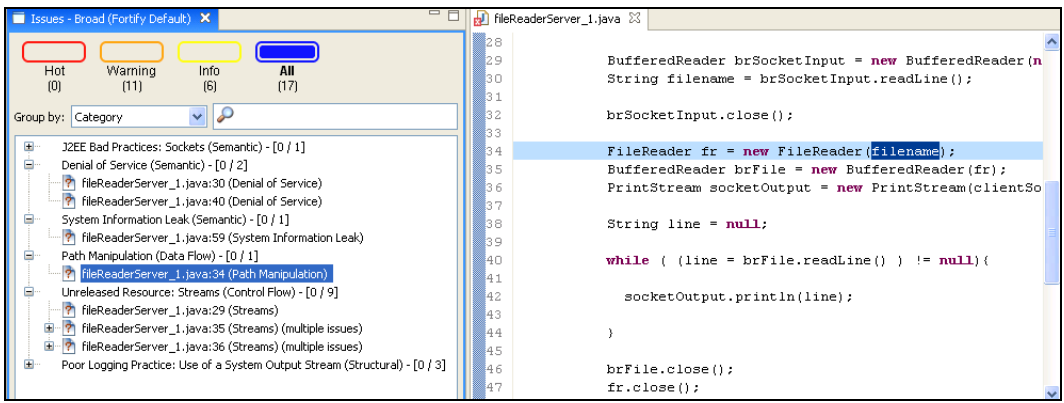

**Figure 7:** Results of the Source Code Analysis of the File Reader Server Program after the Removal of the Resource Injection Vulnerability (fileReaderServer_1.java)

In this section, we present the use of the second approach (i.e. using a level of indirection) to remove the Resource Injection vulnerability (refer to the modified code, especially lines 9 through 25 and 54-56, in Figure 6). The user starting the server program is presented with a list of port numbers to choose from. Each valid port number is presented with a serial number and the user has to choose one among these serial numbers. If the user choice falls outside the valid range of these serial numbers, then the server program terminates printing a simple error message. The limitation is that the user no longer has the liberty to open the server socket at a port number of his choice. This is quite acceptable because often the server sockets are run on specific well-defined port numbers (like HTTP on 80, FTP on 21, etc) and not on arbitrary port numbers, even if the administrator wishes to run the server program on a port number of his choice. Figure 7 presents the results of the source code analysis on the modified connection server socket program (fileReaderServer_1.java) to fix the Resource Injection vulnerability. We have also removed the use of command-line arguments to get inputs and instead use the Scanner class; thus, taking care of the *Leftover debug code* warning.

## 2.2. Path Manipulation Vulnerability

The Path Manipulation vulnerability occurs when user input is directly embedded to the program statements thereby allowing the user to directly control paths employed in file system operations [10]. In our file reader server program, the name or path for the file sent by the client through the socket is received as a String object at the server side, and directly passed onto the FileReader constructor (line 20 in Figure 5). The practice of directly embedding a file name or a path for the file name in the program to access the system resources could be cleverly exploited by a malicious user who may pass an unexpected value for the argument and the consequences of executing the program, especially if it runs with elevated privileges, with that argument may turn out to be fatal. Thus, Path Manipulation vulnerability is a very serious issue and should be definitely not left unattended in a code. Such a vulnerability may enable an attacker to access or modify otherwise protected system resources.

```
7    public static int sanitize(String filename){
8
9        if (filename.indexOf( (int) '/') != -1){
10           System.out.println("invalid argument... You cannot write to a file in other directories..");
11           return -1;
12       }
13
14       if (! filename.endsWith(".txt") ){
15           System.out.println("You can write to only a file with .txt extension...");
16           return -1;
17       }
18
19       return 0;
20
21   }
```

**Figure 8:** Java Code Snippet for the Sanitize Method to Validate the Filename Received through Socket

As suggested in Section 2.1, we propose to use the approach of filtering user inputs using the blacklist/white list approach. It would not be rather advisable to present the list of file names to the client at the remote side – because this would reveal unnecessary system information to a remote user. It would be rather more prudent to let the client to send the name or the path for the file he wants to open, and we validate the input against a set of allowable and non-allowable characters. In this paper, we assume the file requested to be read is located in the same directory from which the server program is run, and that the file is a text file. Hence, the last four characters of the input received through the socket should be ".txt" and nothing else (thus, .txt at the end of the String input constitutes a white list). Also, since the user is not permitted to read a file that is in a directory other than the one in which the server program is running, the input should not have any '/' character (constituting a blacklist) to indicate a path for the file to be read. In this paper, we have implemented the solution of using white list and blacklist through the *sanitize*( ) method, the code for which is illustrated in Figure 8. The modified file server program that calls the sanitize method to validate the filename before opening the file for read is shown in Figure 9. The results of the source code analysis of the modified file reader server program are shown in Figure 10.

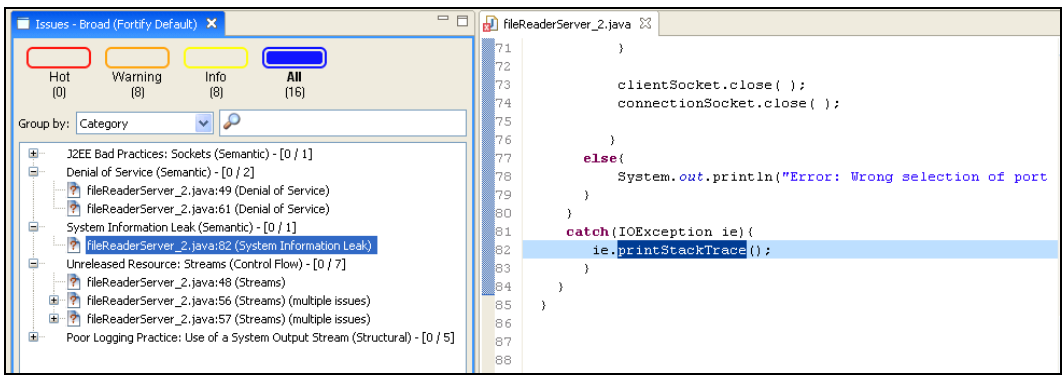

**Figure 9:** Modified File Reader Server Socket Program to Call the Sanitize Method to Validate the Filename before Opening it to Read (fileReaderServer_2.java)

**Figure 10:** Results of the Source Code Analysis of the File Reader Server Program after the Removal of the Path Manipulation Vulnerability (fileReaderServer_2.java)

### 2.3. System Information Leak Vulnerability

The "System Information Leak" vulnerability (a semantic issue) refers to revealing critical system data, program structure including call stack or debugging information that may help an adversary to learn about the software and the system, and form a plan of attack [12]. In our file reader server program (see Figure 6), we observe that in line 82 (as also indicated by the SCA in Audit Workbench Issues panel in Figure 10), the *printStackTrace*( ) method called on the object of the class *IOException* has the vulnerability to leak out sensitive system and program information including its structure and the call stack. While revealing the information about the call stack leading to an exception may be useful for programmers to debug the program and quickly as well as effectively trace out the cause of an error, the *printStackTrace*( ) method needs to be removed from the final program prior to deployment.

A simple fix to this vulnerability is not to reveal much information about the error, and simply state that an error has occurred. The attacker, if he was contemplating to leverage the error information to plan for an attack, would not be able to gain much information from the error message. In this context, we remove the call to the *printStackTrace*( ) method from line 82 and replace it with a print statement just indicating that an error occurred. The modified version of the file reader server socket program is shown in Figure 11 and the results of its source code analysis are shown in Figure 12.

```
24
25      public static void main(){
26         try{
27

76          }
77          else{
78              System.out.println("Error: wrong selection of port number...");
79          }
80      }
81      catch(IOException ie){
82          System.out.println("An error occurred....");
83      }
84      }
85  }
```

**Figure 11:** Modified File Reader Server Program to Remove the System Information Leak Vulnerability (fileReaderServer_3.java)

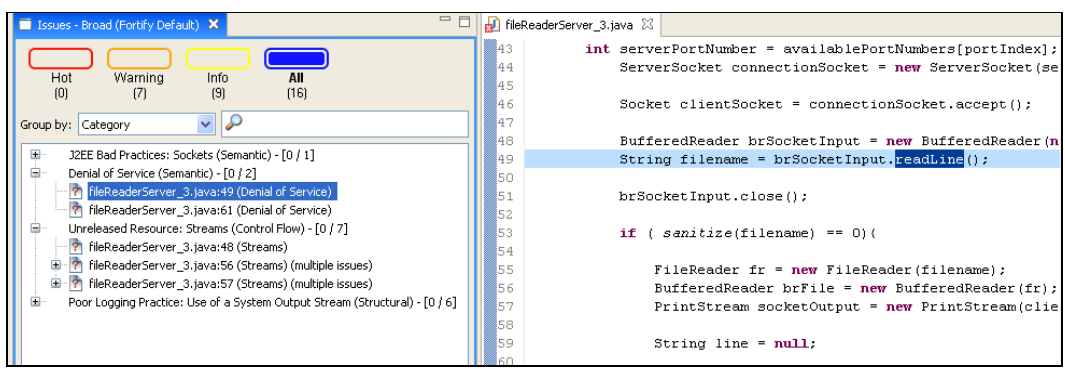

**Figure 12:** Results of the Source Code Analyzer of fileReaderServer_3.java after the Removal of the System Information Leak Vulnerability and Indicating the Presence of the Denial of Service Vulnerability

### 2.4. Denial of Service Vulnerability

A 'Denial of Service' vulnerability (a semantic issue) is the one with which an attacker can cause the program to crash or make it unavailable to legitimate users [10]. Lines 49 and 61 of the file reader server socket program (as indicated in Figure 12) contain the Denial of Service vulnerability, and this is attributed to the use of the readLine( ) method. It is always not a good idea to read a line of characters from a file through a program because the line could contain an arbitrary number of characters, without a prescribed upper limit. An attacker could misuse this and force the program to read an unbounded amount of input as a line through the readLine( ) method. An attacker can take advantage of this code to cause an *OutOfMemoryException* or to consume a large amount of memory so that the program spends more time performing garbage collection or runs out of memory during some subsequent operation.

The solution we suggest is to impose an upper bound on the number of characters that can be read from the file and buffered at a time (i.e., in one single read operation). In this context, we

International Journal of Network Security & Its Applications (IJNSA), Vol.5, No.1, January 2013

suggest to use the *read*( ) method of the *BufferedReader* class that takes three arguments: a character array to which the characters read from the buffer are stored, the starting index in the character array to begin storing characters and the number of characters to be read from the buffer stream. In the context of lines 49 through 54 in the fileReaderServer_4.java program (boxed in Figure 13), we replace the readLine( ) method with a read( ) method to read the name of the file or the pathname. If we do not read sufficient number of characters, then the name of the file stored in the String object *filename* would be incorrect and this could be detected through the current implementation of the sanitize( ) method (Figure 8) itself, as the last four characters of the file has to end in ".txt". In the context of lines 62 through 71 (boxed in Figure 13), there would not be a problem in reading certain number of characters (rather than a line of characters) for every read operation, because – whatever is read is stored as a String and is sent across the socket. In order to preserve the structure of the text, we have to simply use the print( ) method instead of the println( ) method of the PrintStream class. If there is a line break in the text of the file, it would be captured through an embedded '\n' line break character and sent across the socket.

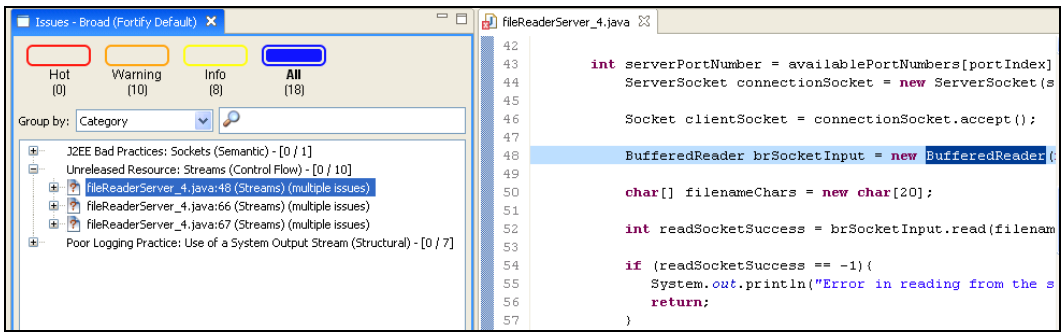

**Figure 13:** Modified Code for the File Reader Server Socket Program to Remove the Denial of Service Vulnerability by Replacing the readLine( ) Method with the read( ) Method (fileReaderServer_4.java)

**Figure 14:** Results of the Source Code Analysis of the File Reader Server Program after Removing the Denial of Service Vulnerability (fileReaderServer_4.java)

In this section, we choose to read 20 characters for each read operation at both the instances and replace the readLine( ) method with the read( ) method accordingly. In the second case, we read every 20 characters from the file, and the last read operation may read less than 20 characters if there are not sufficient characters. The subsequent read will return -1 if no character is read. Our logic (as shown in lines 63-71 of the code in Figure 13) is to check for the return value of the read operation every time and exit the while loop if the return value is -1, indicating the entire file has been read.

Note that the length 20 we used here is arbitrary, and could be even set to 100. The bottom line is there should be a definite upper bound on the number of characters that can be read into the character buffer, and one should not be allowed to read an arbitrary number of characters with no defined upper limit. The modified file reader server socket program is shown in Figure 13 and the results of the source code analysis are shown in Figure 14.

## 2.5. Unreleased Resource Vulnerability

The "Unreleased Resource" vulnerability (a control flow issue) occurs if the program has been coded in such a way that it can potentially fail to release a system resource [11]. In our file reader server socket program, the vulnerability arose due to the use of the *BufferedReader* stream class (lines 48 and 66 of Figure 13) to read the contents from the socket and the text file and the *PrintStream* class to send the contents across the socket to the remote side. Even though we have called the close( ) methods on the objects of the above two stream classes immediately after their use is no longer needed, it may be possible that due to abrupt termination of the program, the close( ) method calls are not executed (also listed in the Issues panel of Figure 14). One possible reason for the program control to skip the execution of the close( ) method calls could be a file read error, which could happen if the name of the file read from the socket is not in the location from which the program is trying to open and read the file. Another reason (which is very unlikely to happen though, given the smaller size of the file) could be that there is no sufficient memory in the system to load the contents of the text file and read them. Similarly, in the case of sending across the socket, there may be an error if the client abruptly closes the socket while the server attempts to transmit them across the socket. Either way, if any such buffer reading or sending errors occur, the program control immediately shifts from the *try* block to the *catch* block and the streams corresponding to the *BufferedReader* and *PrintStream* classes will never be released until the operating system explicitly forces the release of these resources upon the termination of the program. From a security standpoint, if an attacker could sense the presence of Unreleased Resource vulnerability in a program, he can intentionally trigger resource leaks and failures in the operating environment of the program (like making the file unavailable to be read or closing the socket from the remote side, if the client is compromised) to cause a depletion of the resource pool.

The solution we suggest is to add a *finally* { … } block after the *try* {…} *catch* {…} blocks and release all the resources that were used by the code in the corresponding *try* block. Note that in order to do so, the variables associated with the resources have to be declared outside and before the *try* block so that they can be accessed inside the *finally* block. In our case, we have to declare the stream objects of the *BufferedReader* and *PrintStream* classes outside the *try* block and close them explicitly in the *finally* block. The modified code segment (fileReaderServre_5.java) is shown in Figure 15. The results generated from analyzing the fileReaderServer_5.java code with the Source Code Analyzer are shown in Figure 16. Note that in order to close the two *FileReader* and *BufferedReader* streams in lines 66 and 68 of the *finally* {...} block, we have to declare that the main function throws the *IOException* in line 7.

Note that as shown in Figure 15, the reason why we are insisting on including the close( ) method calls on the two stream objects in a *finally* block instead of a *catch* block, even though it is supposed to catch the *IOException*, is that in case a *try* block can generate multiple exceptions – there has to be multiple *catch* blocks for the *try* block, one for each exception, and these *catch*

International Journal of Network Security & Its Applications (IJNSA), Vol.5, No.1, January 2013

blocks have to be listed in the order of increasing scope – i.e., the exception that is the bottommost in the hierarchy of exceptions has to be caught first, followed by exceptions further up in the hierarchy. However, if at run-time, an exception higher up in the hierarchy is generated, the control transfers to the *catch* block of that particular exception, and only the subsequent *catch* blocks are executed, and not the *catch* blocks prior to it. This way, if we had included the close( ) methods in the *catch* block for the *IOException* class and relied on it to be called in case of a file read error, there might be a situation that another *catch* block downstream is called due to the generation of an exception higher up in the exception hierarchy, and the two stream objects would not be released. Thus, in any situation, we do not recommend releasing system resources inside *catch* blocks. The results of the source code analysis of the final version of the file reader server socket program (fileReaderServer_5.java) are shown in Figure 16, after removing all the five main vulnerabilities in the code. The only warnings remaining in Figure 16 are those corresponding to the *poor logging practice* and *J2EE Bad Practices*: *Sockets*, which are not critical to be removed for standard Java programming environments.

```
25    public static void main() throws IOException{
26
27        BufferedReader brSocketInput = null;
28        BufferedReader brFile = null;
29        PrintStream socketOutput = null;
30
31        try{
32

97        }
98        catch(IOException ie){
99            System.out.println("An error occurred....");
100       }
101       finally{
102
103           if (brSocketInput != null)
104               brSocketInput.close();
105           if (brFile != null)
106               brFile.close();
107           if (socketOutput != null)
108               socketOutput.close();
109
110       }
111
112
113       }
114   }
```

**Figure 15:** Modified Code Segment to Remove the Unreleased Resource Vulnerability (fileReaderServer_5.java)

Before we conclude, we also argue that it is not advisable to include a *finalize*( ) method for the particular classes of the objects for which the resources allocated need to be reclaimed. In order for an object's *finalize*( ) method to be invoked, the garbage collector must first of all determine that the object is eligible for garbage collection. However, the garbage collector is not required to run unless the Java Virtual Machine (JVM) is low on memory, and hence there is no guarantee that an object's *finalize*( ) method will be invoked in an expedient fashion. Even if the garbage collector gets to run, all the resources will be reclaimed in a short period of time, and this can lead to "bursty" performance and a reduction in the overall system throughput. Such an effect is more pronounced as the load on the system increases. Also, it is possible for the thread executing the *finalize*( ) method to hang if the resource reclamation operation requires communication over a network or a database connection to complete the operation.

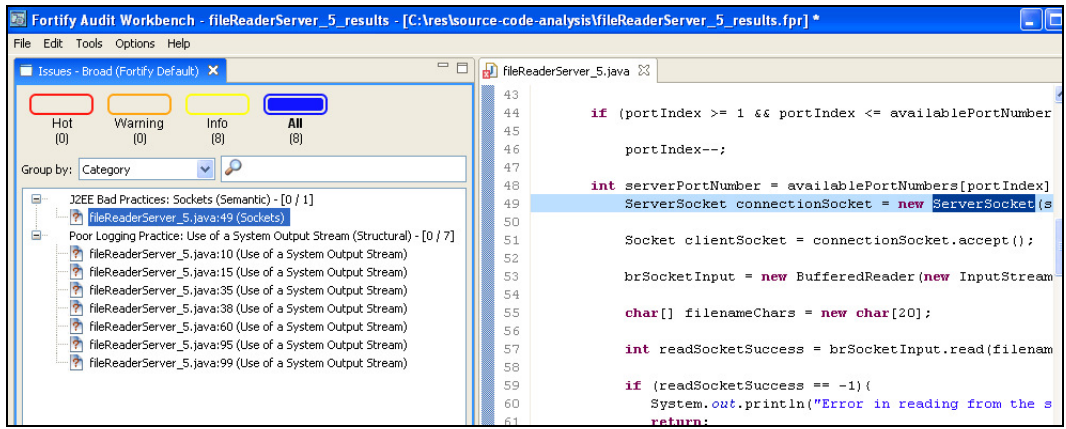

**Figure 16:** Results of Source Code Analysis on the Final Version of the File Reader Server Socket Program with all the Main Vulnerabilities and Warnings Removed (fileReaderServer_5.java)

## 3. A SUMMARIZED OVERVIEW OF THE VULNERABILITIES AND THE SOLUTIONS

In this section, we summarize the seven different vulnerabilities analyzed in the paper and the solutions presented to remove them from the programs.

- **Denial of Service Vulnerability:** The Denial of Service Vulnerability in the file reader program arose because of the use of the *readLine*( ) method that has no upper bound on the number of characters to be read as a line of input from the file. As a result, if an attacker manages to point the FileReader object to a file that has significantly larger number of characters per line, then there is a possibility of overflowing the memory (especially in embedded systems, potentially causing a denial of service attack. The *readLine*( ) method is an atomic operation and the characters that are read as part of a line from the file are not written to the socket until the end-of-line ('\n') character is come across. The solution we suggest is to impose an upper limit on the number of characters that are to be read any time from the file, and do not have the option of letting the user to be able to read an arbitrary number of characters, without any upper limit, in one single read operation.

- **Unreleased Resource Vulnerability:** The Unreleased Resource vulnerability arises when the developer does not include appropriate code that would guarantee the release of the system resources after their intended use (i.e., the resources are no longer needed in the program). In the context of the file writer program, the unreleased resource vulnerability was attributed due to the possibility of the stream objects (the File Reader and Buffered Reader) that were instantiated to read the password text file not being released due to unanticipated errors during the file read operation, and the control flow of the program immediately switching to the *catch* block for the *IOException* (the exception class handling most of the file read errors in Java) and not executing the release statements (the close method calls) for the streams in the *try* block. Once the control exits the *catch* block, the program continues to normally execute the code following the entire *try-catch* block and does not go back to execute the remaining code in the *try* block where the error occurred. Hence, at most care should be taken to release the resources that were allocated inside the *try* block.

However, since there can be more one than *catch* block, for a single *try* block, depending on the specific exceptions that could be generated from the *try* block and need to be caught and processed, it would not be a good idea to include statements pertaining to the release of the allocated system resources in a particular *catch* block. This is because *catch* blocks are supposed to be listed in the increasing order of the scope of the possible exceptions that could be generated from the *try* block, and hence if an exception that is higher up in the hierarchy than an *IOException* is generated before completing the file read operation, then the control would go to a *catch* block that is included somewhere below the *catch* block for the *IOException* class, and the stream resources allocated for the file read operation would not be released at all. To avoid such a scenario, one solution is to include redundant resource release statements in each of the *catch* blocks. This would unnecessarily increase the code size and the memory footprint of the program. An alternate and better solution is to include a *finally* block (optional in Java and need not be used along with a *try-catch* block) following the *catch* blocks and include in it all the statements related to the release of the allocated system resources in the corresponding *try* block. The good news is the Java Virtual Machine will definitely execute the *finally* block after executing one or more *catch* block(s), and will not skip it. This way, the allocated system resources in the *try* block are guaranteed to be released.

- **System Information Leak Vulnerability:** This vulnerability arises when the print statements, in response to unexpected inputs or execution flow, can be used to infer critical information about the program structure, including the sequence of method calls (the call stack). Developers include print statements to facilitate debugging for any erroneous behavior/input. Often such print statements are left in the code even after deployment. Attackers could take advantage of this vulnerability and pass cleverly crafted suitable input values to the program so that it generates the informative error messages. For example, there is a difference between displaying a generic error message like 'Incorrect username or password' compared to a more informative error message like 'Access denied'. In the former case, one cannot easily infer whether the username is wrong or the password is wrong, essentially perplexing the attacker whether or not there exists an account with the particular username passed as input. However, with the 'Access denied' error message, the attacker could easily infer that there exists an account with the attempted username; it is only the password that is invalid. Nevertheless, generic error messages baffle the genuine users of the system and would not be able to infer much from these messages. Thus, there is a tradeoff between the amount of information displayed through the error messages and the ease associated with identifying or debugging the error and fixing the problem.

- **Path Manipulation Vulnerability:** The path manipulation vulnerability arises when the inputs entered by the user are directly embedded into the program and executed, without validating the correctness and appropriateness of the input values. The vulnerability becomes more critical if the user values are directly embedded into path statements that read a critical system resource (say a file) and the program executes them with elevated privileges (higher than that of the user who passed the input). We present two solution approaches to handle this vulnerability (presenting a list of probable inputs only from which the user can choose his choice or sanitizing/filtering the user input by validating it with a blacklist of non-allowable characters and a white list of allowable characters). We adopt the second approach and present a sanitizer code in Java that scans the user input for the path/name of the log file with respect to a blacklist of non-allowable characters and a white list of allowable characters.

- **Resource Injection Vulnerability:** Like the Path Manipulation vulnerability, the Resource Injection vulnerability occurs due to the embedding of user entered inputs to a program so that the attacker gets direct access to a resource identifier (like ports) that would have been otherwise inaccessible. In the connection-oriented server socket program, we illustrate the incidence of the vulnerability by letting a user to open a server socket at the port number of his choice. Even though server programs are typically started by the system administrator, if the system is under attack – this feature could be misused and an attacker can start the server program at a port number of his choice, typically by passing port numbers that are reserved for other regular/critical system processes. We remove this vulnerability by presenting the user a list of port numbers to choose from, and the user has no option other than choosing one among those in the list. Though this option reveals the list of available port numbers, the user is constrained to choose only from this list.

## 4. CONCLUSIONS AND FUTURE WORK

Software security is a rapidly growing field and is most sought after in both industry and academics. With the development of automated tools such as Fortify Source Code Analyzer, it becomes more tenable for a software developer to fix, in-house, the vulnerabilities associated with the software prior to its release and reduce the number of patches that need to be applied to the software after its release. In this paper, we have discussed the use of an automated tool called the Source Code Analyzer (SCA), developed by Fortify, Inc., and illustrated the use of its command line and graphical user interface (Audit Workbench) options to present and analyze the vulnerabilities identified in a software program. The SCA could be used in a variety of platforms and several object-oriented programming languages. We present an exhaustive case study of a file reader server socket program, developed in Java, which looks fine at the outset; but is analyzed to contain critical vulnerabilities that could have serious impacts when exploited.

The five different vulnerabilities we have studied in this research are: Resource Injection vulnerability, Path Manipulation vulnerability, System Information Leak vulnerability, Denial of Service vulnerability, and Unreleased Resource vulnerability in the context of streams. We discussed the reasons these vulnerabilities appeared in the code and how they could be exploited if left unattended and the consequences of an attack. We have provided detailed solutions to efficiently and effectively remove each of these vulnerabilities, presented the appropriate code snippets and the results of source code analysis when the vulnerabilities are fixed one after the other. The tradeoffs incurred due to the incorporation of appropriate solutions to fix these vulnerabilities are the increase in code size and decrease in the comfort level for a naïve authentic user who could face some initial technical difficulties in getting the program to run as desired. With generic error messages that are not so detailed, an authentic (but relatively unfamiliar) user ends up spending more time to run the system as desired. The original file reader server program had 43 lines of code, and the final version of the program (fileReaderServer_5.java) contains 114 lines – thus, an increase in the size of the code by a factor of about 2.65 (i.e., 165% increase). However, the increase in code size is worth because even if one the above 5 vulnerabilities is exploited by an attacker, it could be catastrophic for the entire network hosting the server.

As part of future work, we plan to conduct exhaustive source code analysis on network socket programs developed in C/C++, for Windows and Linux platforms, and analyze their impacts and develop effective solutions to fix (i.e., completely remove or mitigate the effects as much as possible) the characteristic vulnerabilities identified for the specific platform/ programming language. Even though the code snippets provided as solutions to remove the various software security vulnerabilities discussed in this paper are written in Java, the solutions proposed and implemented here for each of the vulnerabilities are more generic and can be appropriately modified and applied in other programming language environments.

## ACKNOWLEDGMENTS

The work leading to this paper is partly funded through the U. S. National Science Foundation (NSF) CCLI/TUES grant (DUE-0941959) on "Incorporating Systems Security and Software Security in Senior Projects." The views and conclusions contained in this document are those of the author and should not be interpreted as necessarily representing the official policies, either expressed or implied, of the funding agency.

## REFERENCES

[1] B. Chess, and J. West, *Secure Programming with Static Analysis*, Addison Wesley, 1st Edition, Boston, MA, USA, 2008.

[2] M. R. Stytz, and S. B. Banks, "Dynamic Software Security Testing," *IEEE Security and Privacy*, vol. 4, no. 3, pp. 77-79, 2006.

[3] D. Baca, "Static Code Analysis to Detect Software Security Vulnerabilities – Does Experience Matter?," *Proceedings of the IEEE International Conference on Availability, Reliability and Security*, pp. 804-810, 2009.

[4] P. R. Caseley, and M. J. Hadley, "Assessing the Effectiveness of Static Code Analysis," *Proceedings of the 1st Institution of Engineering and Technology International Conference on System Safety*, pp. 227-237, 2006.

[5] I. A. Tondel, M. G. Jaatun and J. Jensen, "Learning from Software Security Testing," *Proceedings of the International Conference on Software Testing Verification and Validation Workshop*, pp. 286-294, 2008.

[6] H. Mcheick, H. Dhiab, M. Dbouk and R. Mcheik, "Detecting Type Errors and Secure Coding in C/C++ Applications," *Proceedings of the IEEE/ACS International Conference on Computer Systems and Applications*, pp. 1-9, 2010.

[7] M. Mantere, I. Uusitalo and J. Roning, "Comparison of Static Code Analysis Tools," *Proceedings of the 3rd International Conference on Emerging Security Information, Systems and Technologies*, pp. 15-22, 2009.

[8] J. Novak, A. Krajnc and R. Zontar, "Taxonomy of Static Code Analysis Tools," *Proceedings of the 33rd IEEE International Conference on Information and Communication Technology, Electronics and Microelectronics*, pp. 418-422, 2010.

[9] https://www.fortify.com/products/hpfssc/source-code-analyzer.html, last accessed: July 2, 2012.

[10] M. G. Graff, and K. R. Van Wyk, *Secure Coding: Principles and Practices*, O'Reilly Media, Sebastopol, CA, USA, 2003.

[11] M. Howard, D. Leblanc, and J. Viega, *24 Deadly Sins of Software Security*: Programming Flaws and How to Fix them, McGraw-Hill, New York City, NY, USA, 2009.

[12] J. A. Whittaker, *How to Break Software*, Addison-Wesley, Boston, MA, USA, 2002.

**Author**

Dr. Natarajan Meghanathan is a tenured Associate Professor of Computer Science at Jackson State University. He graduated with MS and PhD degrees in Computer Science from Auburn University and The University of Texas at Dallas respectively. He has authored more than 140 peer-reviewed publications. He has received federal grants from the U. S. National Science Foundation, Army Research Lab and Air Force Research Lab. He is serving in the editorial board of several international journals and in the organization/program committees of several international conferences. His research interests are: Wireless Ad hoc Networks and Sensor Networks, Network Security and Software Security, Graph Theory, Computational Biology and Cloud Computing. 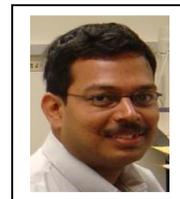